\documentclass{phbauth}

\usepackage{epsfig}

\begin{document}

\begin{frontmatter}
\title{
On direction of spontaneous magnetization in a "cubic" ferromagnet}

\author{A.V.Kovalev\thanksref{thank1}}
\and\
\author{L.A. Akselrod}

\address{Petersburg Nuclear Physics Institute, 188300, Gatchina, Russia}

\thanks[thank1]{Corresponding author.\\
E-mail: kovalev@hep486.pnpi.spb.ru }

\begin{abstract}

The magnetic properties of anisotropic films have been studied using 3D-neutron polarization analysis. The experimental facts refer to essential
distinction of the sample states, magnetized in opposite directions. For
an explanation of asymmetrical effects the model is offered, in which the
fundamental theoretical principles of structural phase transitions are
used. \end{abstract}


\end{frontmatter}

\section{Introduction}
\begin{figure}
\centerline{\epsfxsize=7cm \epsfysize=7cm \epsfbox{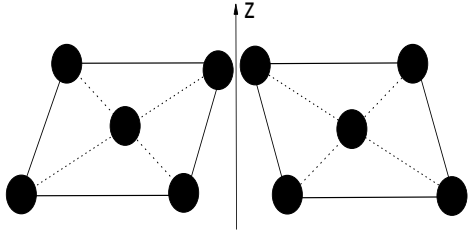}}
\caption{
Twins of the monoclinic phase in a (110) plane.\label{fig:fig1}}
\end{figure}

The central point of the present work is the hypothesis about the quite certain
direction of a vector of the spontaneous magnetization ${\bf M}_s$ in a single crystal of a magnetic
phase $G_m$ \cite{1}. In the case under study the orientation ${\bf
M}_s\|$[110] takes place. Hence, the phase $G_m$ should have the
monoclinic symmetry $2/m$ (Neumann principle), and $bcc$ cell should be
deformed along the directions of the types [001], [110] and [111]. In a
crystal lattice of a  paramagnetic phase $G_p$ after structural phase
transition $G_p\to G_m$ the twins of a magnetic phase will be formed.
The number of various orientations n of such twins is equal to the
number of equivalent directions [hkl] of a phase $G_p$, which the
direction of ${\bf M}_s$ coincides with. For example, for a cubic phase
and with ${\bf M}_s\|$[110] this number is n = 12.  The component of
the spontaneous deformation of a bcc cell along a direction of the type
[111] results in formation of the twins (fig.~\ref{fig:fig1}). The atoms
are located in the corners and centers of a parallelograms disposed in
a (110) plane. Such twins are transformed into each other by a halfturn
around  ${\bf z}$-axis. Thus, the directions of all magnetic moments of
atoms (spins) change also. The following assumption, therefore, seems
quite natural: the spontaneous magnetization in the considered twins
has opposite directions. Thus, the problem arises of definition of a
"correct" direction of ${\bf M}_s$. The study of the possibilities of
vector analysis of polarized neutrons for the solution of this problem
was the initial purpose of the present work.

\section{The samples and the method of measurements}
\begin{figure}
\centerline{\epsfxsize=7cm \epsfysize= 7cm \epsfbox{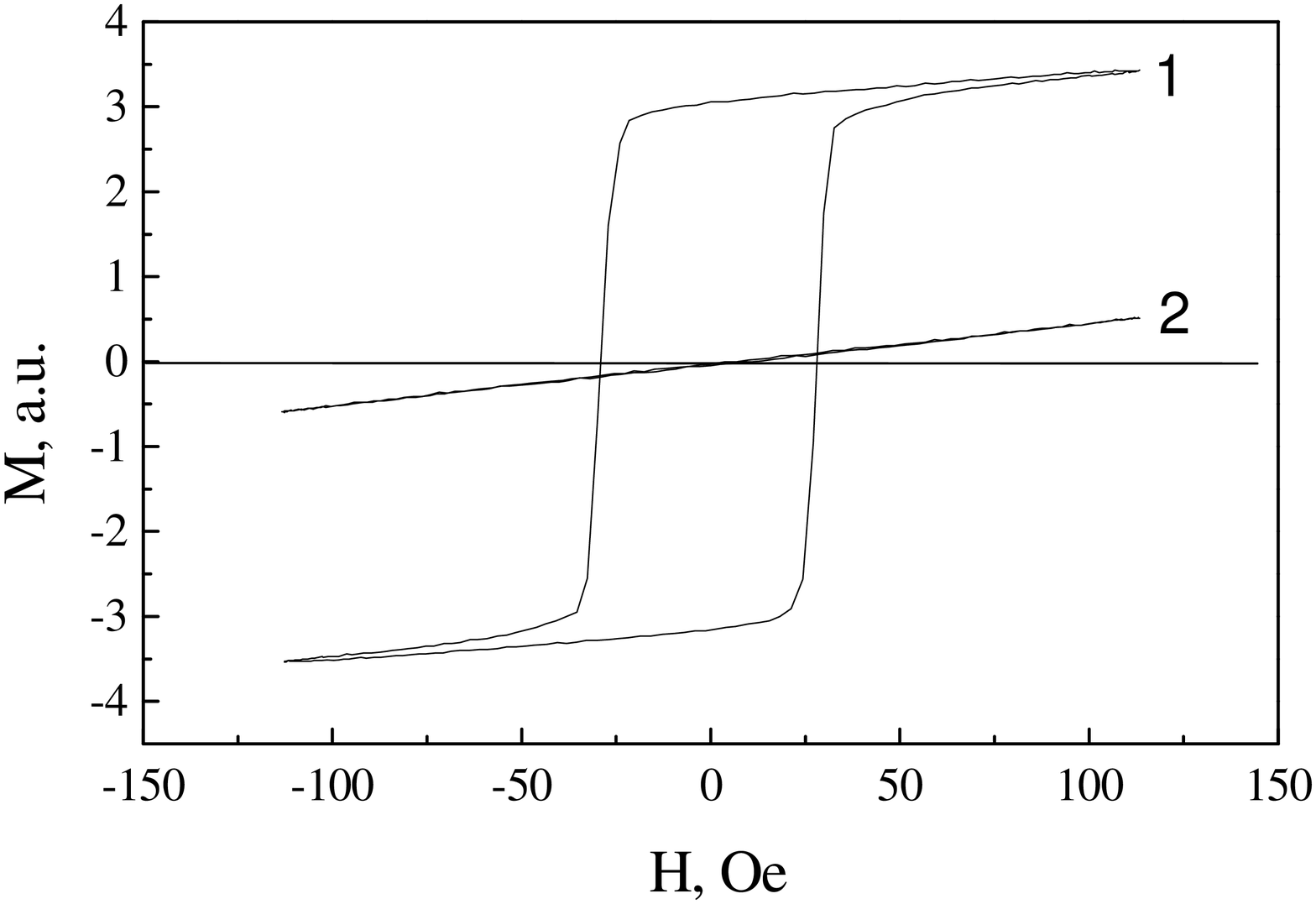}}
\caption{
Gysteresis loops for the easy (1) and hard (2)
directions.\label{fig:fig2}}
\end{figure}

The choice of samples in our case plays a crucial role. The optimum way would be to
use a single crystal of a monoclinic phase. We worked with the more complex system:
anisotropic polycrystalline Co-Fe films, in which uniaxial texture of the twins is formed
by the asymmetry of the fluxes of deposited particles \cite{2}. In
particular, the ideal model of this texture is given on  fig.1. The
films 3 $\mu$m thick were manufactured by magnetron sputtering on glass
substrates. During the films deposition the metastable states of the
crystalline structure are formed which stabilize themselves at annealing temperatures of
the order 150$^\circ$C \cite{1}.   On hysteresis loops
(fig.~\ref{fig:fig2}), measured in a magnetic field of the frequency 50 Hz,
the strong distinction of magnetic properties along easy and hard axes
(EA and HA) is observed. For the initial treatment of the films the permanent magnet with
$H=1500$ Oe is used.

The diagram of the setup for 3D analysis of polarized neutrons and the general
concept of measurements are given in \cite{3}, therefore, here we shall
note only the basic moments regarding the  present work. The neutron
beam directed along the {\bf x}-axis falls  on a film, initially lying
in yz-plane ({\bf z} - vertical direction) and  EA  coincides with {\bf
z}-axis.  The measurements are performed at constant $\vartheta$ for a
number of orientations as the sample is rotated around  $\bf z$-axis up
to $\vartheta = 80^\circ$. Because the coercive force $H_c$ for our
films is about 40 Oe, the external magnetic field $H < 0.05$ Oe in the
region of the sample can be neglected. The primary polarization ${\bf
P}_0$ is fixed  along the coordinate axes in turn. After passing the
sample for each directions of the primary polarization  three
components $P_{ij}$ of the vectors ${\bf P}_i(P_x, P_y, P_z)$ are
measured. In model calculations the normalized values $P_{ij}$ are
used.

With homogeneous distribution of a magnetic field $\bf B$ in a sample the magnitude of B
is calculated according to the expressions:

\begin{eqnarray}
& P_{xx} = P_{yy} = cos\psi,&\nonumber\\& P_{xy} = - P_{yx}  =
sin\psi,&\\ &\psi(radn) = 0.106\cdot B\cdot(G)\cdot L(cm),&\nonumber
\end{eqnarray}

where $L = d/cos\vartheta$ is the effective thickness of a sample in the direction of a neutron beam.
In this case there is no depolarization of the beam at all, i.e. $M_i \equiv \mid {\bf P}_i \mid = 1$. The main
results of the studies of asymmetrical effects are presented below.

\section{Results and discussions}
\begin{figure}
\centerline{\epsfxsize=7cm\epsfysize=7cm\epsfbox{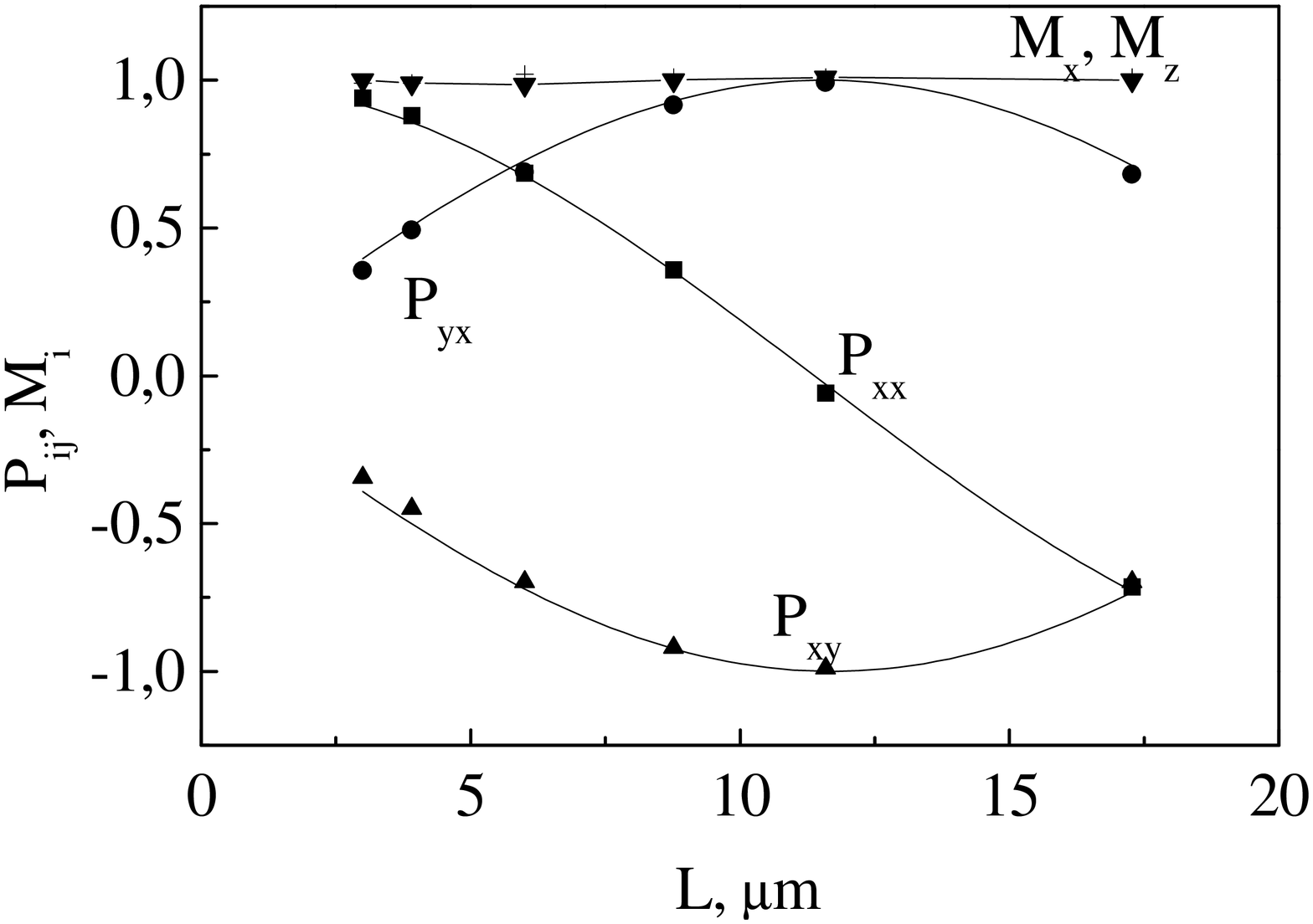}}
\caption{
Uniform magnetization. Some components $P_{ij}$ of the polarization vectors ${\bf P}_i$ and
the absolute values $M_i=\mid {\bf P}_i\mid$ as a function of the
neutron path L in the film.\label{fig:fig3}} \end{figure}

The characteristic dependencies $P_{ij}(L)$ and $M_i(L)$ for the uniformly magnetized
samples are presented in fig.~\ref{fig:fig3}. The initial state of the
sputtered films appears to be uniformly magnetized also. The possible
spatial distribution of vectors $\bf B$ around EA- direction results in
the small distinctions of B values, calculated from the different
components $P_{ij}(L)$. For example,  for an initial state they are
$B(P_{xx})= 11730(170) G$ and $B(P_{xy})=12220(210) G$. Therefore , the
average $<B>$ will be employed below.

After the annealing of the sample at $t\approx 200^\circ C$ there is a small increase of the
induction $<B>$ up to 12840(200) G only. After the sample is being dragged between the
poles of a magnet, in such a way that ${\bf H}\parallel \bf B$, $<B>$ does not change. But after the similar
operation with the opposite direction $\bf H$ the change of the direction $\bf B$ and its reduction to
$<B> = 10400(300) G$ has been observed, that leads to a conclusion on the distinction of
the states of the film magnetized in opposite directions.

\begin{figure}
\centerline{\epsfxsize=7cm\epsfysize=7cm\epsfbox{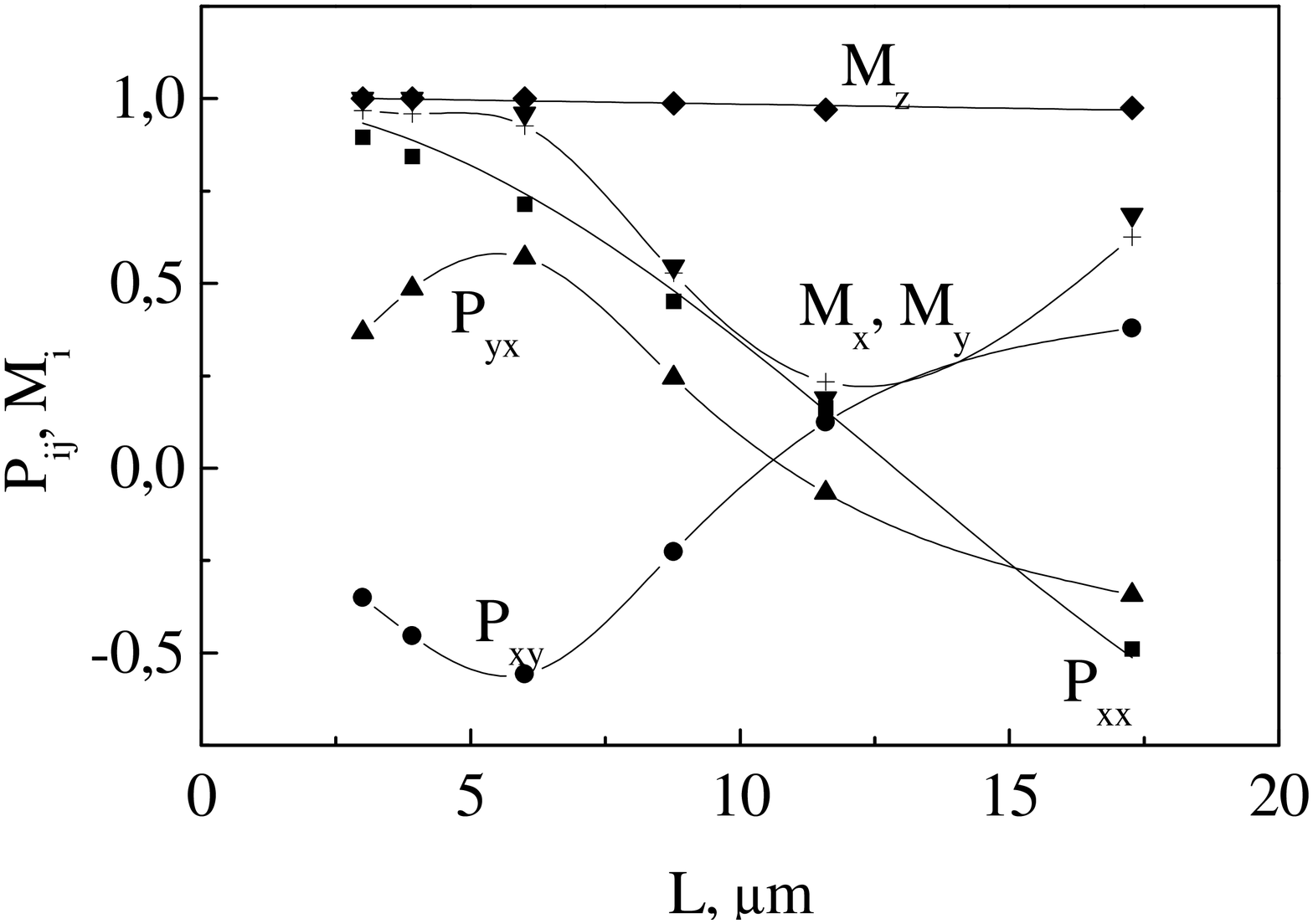}}
\caption{
Multi-domain structure. Some components $P_{ij}$ of the polarization vectors ${\bf P}_i$ and
the absolute values $M_i=\mid {\bf P}_i\mid$ as a function of the neutron path L in the film.
\label{fig:fig4}} \end{figure}

The qualitative effects have been observed in another measurements. The magnetic
field was applied on the film at the angle $\varphi$ to  EA-direction. One can assume, that at
$\varphi=90^\circ$ and after extracting of the film from the magnet a multi-domain structure is
generated, that can be registered easily by the occurence of strong depolarization of the
neutron beam (fig.~\ref{fig:fig4}). It turned out, that  for
$\varphi=90^\circ$ one can obtain such a state for one direction of the
initial magnetization only and after a number of trials. Usually, after
such operation either the initial state of magnetization restores or it
turns by $180^\circ$.  For the opposite initial magnetization the
change of the $\bf B$ direction occurs in an interval of angles from
$100^\circ$ up to $110^\circ$. Besides, the values
$<B_1(\varphi=100^\circ)>=13120(240)G$ and
$<B2(\varphi=110^\circ)>=10600(280)G$ coincide with the values measured
in the first experiment.  These facts indicate the different
stabilities of the magnetized film states.

The asymmetrical effects are in contrast to the results of the magnetic measurements
(fig. 2): the residual magnetization $M_r$ is equal to $-M_r$. So far, the reason of these
contradictions has not been properly understood. But on the basis of the considerable
evidences one can put forward a tentative model for an explanation of the asymmetrical
effects.

In the films, under certain crystallization conditions, the different concentrations of
the twins of the monoclinic phase form. It is reasonable to call the "correct" direction of
${\bf B \parallel M}_s$, as the ground state of the magnetized crystal. The change of the direction ${\bf B\to -B}$
in an external magnetic field can result in  different states of the crystal structure of the
twin because of the essential difference between the spontaneous deformation and
magnetostriction (MS). The MS distortion can change, in principle, the orientation of the
twin that is equivalent to its turn. Then the two states of the film, which differ by the
opposite magnetizations, will have the same energy and in this case the asymmetrical
effects should not be observed. But the external magnetic field can result in the small
distortions of parallelograms represented in fig.1. The reset of the  magnetization to the
"normal" condition is prevented by the crystal anisotropy and the metastable state retains
without the external field.

It is worth to note that the different stability of the opposite magnetized states can be
connected with the elastic strains. But it is difficult to explain the measured difference
between $<B_1>$ and $<B_2>$ by this reason.

\section{Acknowledgements}
The authors are deeply grateful to B.G.~Peskov and A.V.~Zaitsev for preparation of the
films, to V.V.~Deriglazov, G.P.~Gordeev, A.I.~Okorokov and V.N.~Zabenkin for kind
assistance.

This work was supported by Russian Fund for Basic Research (Grant No.00-15-
96814) and Russia State Programme "Neutron Research of Matter".

\end{document}